\pdfoutput=1
\documentclass[authoryear,final,3p,times]{elsarticle}
\usepackage{amssymb}
\usepackage{amsthm}
\usepackage{amsmath}
\usepackage{bm}
\usepackage{array}
\usepackage{longtable}
\usepackage{setspace}
\usepackage{multirow}
\usepackage{siunitx}
\usepackage{threeparttable}
\usepackage{verbatim}
\usepackage[justification=centering]{caption}
\usepackage{lineno}
\begin{document}
\begin{spacing}{1.0}
\begin{frontmatter}
%\journal{Chemical Engineering Science}
\title{Physics-informed dynamic mode decomposition for short-term and long-term prediction of gas-solid flows}
\author[label1,label2]{Dandan Li}
\author[label1,label2,label3]{Bidan Zhao}
\author[label1,label2]{Shuai Lu}
\author[label1,label2,label3]{Junwu Wang\corref{cor1}}
\cortext[cor1]{Corresponding author}
\ead{jwwang@ipe.ac.cn}
\address[label1]{State Key Laboratory of Multiphase Complex Systems, Institute of Process Engineering, Chinese Academy of Sciences, P. O. Box 353, Beijing 100190, P. R. China}
\address[label2]{School of Chemical Engineering, University of Chinese Academy of Sciences, Beijing 100049, P. R. China}
\address[label3]{Innovation Academy for Green Manufacture, Chinese Academy of Sciences, Beijing 100190, P. R. China}

\begin{abstract}
Integration of physics principles with data-driven methods has attracted great attention in recent few years. In this study, a physics-informed dynamic mode decomposition (piDMD) method, where the mass conservation law is integrated with a purely data-driven DMD method, is developed for fast prediction of the spatiotemporal dynamics of solid volume fraction distribution in bubbling fluidized beds.
Assessment of the prediction ability using both piDMD and DMD is performed using the CFD-DEM results as the benchmark: Both DMD and piDMD can predict the short-term behaviour of solid volume fraction reasonably well, but piDMD outperforms the DMD in both qualitative and quantitative comparisons; With respect to their long-term ability, the piDMD-based prediction of the instantaneous solid volume fraction distributions is qualitatively correct although the accuracy needs to be improved, and the predicted time-averaged radial and axial profiles are satisfactory; Whereas the DMD-based prediction of instantaneous snapshots and time-averaged results is completely nonphysical. Present study provides a fast and relatively accurate method for predicting the hydrodynamics of gas-solid flows.
\end{abstract}
\begin{keyword}
Physics-informed dynamic mode decomposition; Gas-solid system; Multiphase flow; Fluidization; Data-driven method; Particle technology
\end{keyword}
\end{frontmatter}
%\linenumbers

\section{Introduction}\label{s1}
Gas-solid flows are widely encountered in industry and in nature \citep{wang2020continuum}. For example, fluidization technology is one of the most important and widely-used technologies in industrial processes, such as the fluid catalytic cracking, the calcination of ores, the polymerization of olefins, the combustion of coal, municipal waste, and biomass \citep{fan1996summary}. In recent decades, computational multiphase fluid dynamics has been a pivotal and fast-growing research focus of chemical engineering field \citep{van2006multiscale,ge2019multiscale,zhang2023numerical}, one of its major expectations is that the design, optimized operation, scale-up, and dynamic control of chemical reactors can be carried out digitally, without laborious and expensive experiments. For instance, the concept of virtual process engineering have been proposed in recent years to replace the traditional trial-and-error strategy of reactor design and scale-up \citep{ge2011meso,ge2019mesoscience}, which has put forward an extremely high demand on the accuracy and speed of computational fluid dynamics (CFD) simulations. Clearly, the dream of chemical engineers is fully in line with the broader effort in the development of the most promising and enabling digital twin technology for realizing smart manufacturing and Industry 4.0 \citep{tao2018digital,jones2020characterising,baranidharan2022potentials}.

One of the critical steps towards the realization of digital twin of chemical reactors is the capability of simulating and predicting chemical reactors accurately and quickly, including the fluidized bed reactor studied in present article. Various approaches have been developed for fast simulation and/or prediction of gas-solid flows in fluidized bed reactors, such as recurrence CFD \citep{lichtenegger2016recurrence,lichtenegger2018local,dabbagh2021fast}, deep learning \citep{zhang2020modeling,bazai2021using,faridi2023spatio,qin2023deep,wen2023rapid}, GPU-based computation \citep{xu2011quasi,xu2022discrete} and reduced-order model \citep{yuan2005reduced,yu2015dynamic,zhong2020cfd,yu2021coupling,li2022development,hajisharifi2023non}.
For example, for the prediction of solid phase volume fraction distribution in  gas-solid fluidized beds,
%\cite{lichtenegger2018local} suggested that the complicated long-term gas-solid bubbling fluidized bed's motion could be approximated with information from short-term studies by using recurrence CFD (rCFD) simulations, and the higher the system’s complexity, the longer it takes to observe recurrences with a much higher degree of similarity than those after the initial phase;The above methods of the prediction of solid phase volume fraction in the gas-solid fluidized bed have a complex prediction process and limited prediction time.
\cite{bazai2021using} trained a convolutional neural network (CNN) using CFD data, the trained CNN model predicts the next time-step solid volume fraction distribution quicker than the CFD simulation, however, the long-term prediction using the proposed CNN requires constant inputs of CFD data in order not to lose its accuracy;
\cite{qin2023deep} presented a new approach of deep learning for voidage prediction.
 % (DeepVP)accelerating the 2D voidage distribution prediction for a gas-solid fluidized bed at a steady state, but it did not discussed the ability of DeepVP to predict the instantaneous results of the flow field, which is an essential part of the flow prediction, and DeepVP required much time to build a dataset and training model.

With the successful application of data-driven methods in extraction of coherent structures and reconstruction of flow field for gas-solid systems \citep{cizmas2003proper,haghgoo2019analyzing,higham2020using,li2022pod,li2023data},
recent developments in data-driven methods have found potentials in simulating complex fluid motion for computational acceleration and fast prediction \citep{curtis2021dynamic,ghadami2022data,baddoo2023physics}.
In order to predict gas-solid fluidized beds efficiently, this study develops a physics-informed dynamic mode decomposition (piDMD) method, which combines the data-driven approach with the mass conservation law, to predict the solid phase volume fraction of a bubbling fluidized bed. The remaining sections of the article are: Section \ref{s2} introduces the piDMD method; Section \ref{s3} is an introduction to the CFD-DEM method; Section \ref{s4} analyses and discusses the results; and Section \ref{s7} is a summary and outlook of present article.

\section{Physics-informed Dynamic mode decomposition}\label{s2}
\begin{figure}
\centerline{\includegraphics[width=1\textwidth]{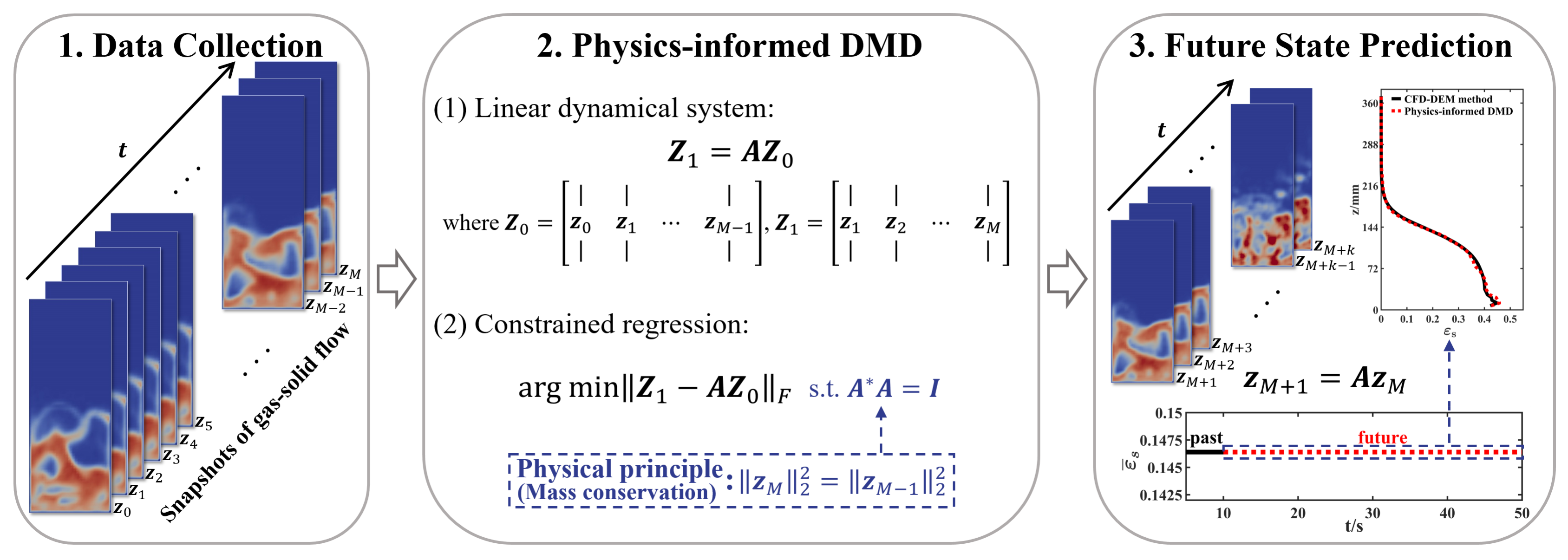}}
\caption{Illustration of prediction of dense gas-solid flow using physics-informed dynamic mode decomposition (piDMD). ${I}$ is the identity matrix, the operators $(\cdot)^{*}$, ${\left\| \cdot \right\|_F}$, and ${\left\| \cdot \right\|_2}$ denote the conjugate transpose operation, the Frobenius norm, and the 2-norm.}
\label{fig1}
\end{figure}

The DMD method \citep{rowley2009spectral,schmid2010dynamic,schmid2022dynamic} is a powerful mathematical tool, which uses the infinite dimensional linear Koopman operator to analyze the nonlinear behaviour in finite dimensions \citep{mezic2013analysis,brunton2021modern}. Collecting data from a series of time snapshots of the flow field is the first step of DMD. As illustrated on the left side of Fig.\ref{fig1}, a snapshot obtained from numerical simulation corresponds to a vector in data analysis.
The specific task of data collection is to arrange the data of each snapshot into a vector according to a fixed sequence of spatial location (grid position), the resulting vectors are then further sorted to obtain the snapshot matrix according to time series.
Assuming that $\mathbf{z}_i$ is the vector composed of a snapshot at the $i$th time, and vectors at different time steps are ordered chronologically to form two snapshot matrices $\mathbf{Z}_0$ and $\mathbf{Z}_1$ that differ by unit time step as shown below,
\begin{equation}
\begin{split}
%&{\mathbf{Z}_0} = [\begin{array}{*{20}{c}} {{\mathbf{z}_0}}&{{\mathbf{z}_1}}&{\cdots}&{{\mathbf{z}_{M-1}}} \end{array}], \\
&{{\bf{Z}}_0} = \left[ {\begin{array}{*{20}{c}}
{\begin{array}{*{20}{c}}
{\rm{|}}\\
{{{\bf{z}}_0}}\\
{\rm{|}}
\end{array}}&{\begin{array}{*{20}{c}}
{\rm{|}}\\
{{{\bf{z}}_1}}\\
{\rm{|}}
\end{array}}&{\begin{array}{*{20}{c}}
{\rm{|}}\\
 \cdots \\
{\rm{|}}
\end{array}}&{\begin{array}{*{20}{c}}
{\rm{|}}\\
{{{\bf{z}}_{M - 1}}}\\
{\rm{|}}
\end{array}}
\end{array}} \right], \\
%&{\mathbf{Z}_1} = [\begin{array}{*{20}{c}} {{\mathbf{z}_1}}&{{\mathbf{z}_2}}&{\cdots}&{{\mathbf{z}_{M}}} \end{array}], \\
&{{\bf{Z}}_1} = \left[ {\begin{array}{*{20}{c}}
{\begin{array}{*{20}{c}}
{\rm{|}}\\
{{{\bf{z}}_1}}\\
{\rm{|}}
\end{array}}&{\begin{array}{*{20}{c}}
{\rm{|}}\\
{{{\bf{z}}_2}}\\
{\rm{|}}
\end{array}}&{\begin{array}{*{20}{c}}
{\rm{|}}\\
 \cdots \\
{\rm{|}}
\end{array}}&{\begin{array}{*{20}{c}}
{\rm{|}}\\
{{{\bf{z}}_M}}\\
{\rm{|}}
\end{array}}
\end{array}} \right], \\
&{\mathbf{Z}_0},\;{\mathbf{Z}_1}\in {R^{N \times M}},
\end{split}
\label{A1}
\end{equation}
where $M + 1$ is the total number of snapshots, and $N$ is the number of spatial locations or the number of computational cells.
The underlying linearity assumption of the DMD method is
\begin{equation}
{\mathbf{z}_{i + 1}} = \mathbf{A} {\mathbf{z}_i},
\label{A2}
\end{equation}

\noindent whose rollout to a sequential set of snapshots is ${\mathbf{Z}}_1 = \mathbf{A} {\mathbf{Z}}_0$. DMD aims to make $\mathbf{A}$ an optimal low-rank approximation matrix as a DMD operator to approximate the dynamical process in the nonlinear dynamical system.
The DMD regression, whose minimization is phrased in the Frobenius norm, is formulated as
\begin{equation}
\mathop {\arg \min }\limits_{{\rm{rank}}\left( {\bf{A}} \right) = r} {\left\| {{{\bf{Z}}_1} - {\bf{A}}{{\bf{Z}}_0}} \right\|_F}.
\label{A3}
\end{equation}
Finding the optimal $\mathbf{A}$ that satisfies Eq. \ref{A3} is the key to the DMD process. The standard DMD algorithm for prediction is based on the singular value decomposition (SVD) \citep{stewart1993early} to solve Eq. \ref{A3}.
SVD decomposes any matrix into the multiplication of three matrices, i.e. an orthogonal matrix $\bf{U}$ composed of left singular vectors, a diagonal matrix $\bf{\Sigma }$, and an orthogonal matrix ${\bf{V}}^*$ composed of right singular vectors, an example is ${{\bf{Z}}_0}={\bf{U\Sigma }}{{\bf{V}}^*}$. SVD is obtained by conventional matrix operations and is easy to implement in the programming language. The contribution of SVD to the DMD-based prediction is to transform the matrix ${\bf{Z}}_0$ that cannot be inverted into three square matrices ($\bf{U}$, ${\bf{\Sigma}}$, and ${\bf{V}}$) that can be inverted.
The specific process of DMD-based prediction of solid phase volume fraction is as follows \citep{schmid2010dynamic,tu2013dynamic}:

\noindent ${1.}$${\,}$Obtain two snapshot matrices $\mathbf{Z}_0$ and $\mathbf{Z}_1$ from Eq. \ref{A1}.

\noindent ${2.}$${\,}$Compute the SVD of ${{\bf{Z}}_0}$,
\begin{equation}
{{\bf{Z}}_0}={\bf{U\Sigma }}{{\bf{V}}^*}.
\label{A31}
\end{equation}

\noindent ${3.}$${\,}$Calculate the high-dimensional mapping matrix $\mathbf{A}$ using the following equation,
\begin{equation}
{\bf{A}} = {{\bf{Z}}_1}{{{\bf{Z}}_0}^{\dagger}} = {{\bf{Z}}_1} {\bf{V}} {{\bf{\Sigma}}^{-1}} {{\bf{U}}^*},
\label{A4}
\end{equation}
where $(\cdot)^{\dagger}$ represents a pseudoinverse operation and $(\cdot)^{-1}$ represents an inverse operation.

\noindent ${4.}$${\,}$By using the obtained $\mathbf{A}$ and the solid volume fraction of the flow field at the $j$th time, which is expressed as $\mathbf{z}_{j}$, the solid volume fraction of the flow field at the $j+1$th time, which is expressed as $\mathbf{z}_{j + 1}$, can then be predicted using the formula as
\begin{equation}
{\mathbf{z}_{j + 1}} = \mathbf{A} {\mathbf{z}_j}.
\label{A5}
\end{equation}
Then the DMD-based prediction of solid volume fraction at the next moment only requires repeating step $4$.

The DMD method is a purely data-driven method, which cannot automatically satisfy the known physical laws of studied systems. In order to remedy this deficiency, physics-informed dynamic mode decomposition (piDMD) has been proposed recently \citep{baddoo2023physics}. The piDMD method is based on the purely data-driven DMD method with the addition of the underlying physical laws of the system as constraints, so systems with different physical laws correspond to different piDMD methods. Via analyzing the system characteristics of gas-solid bubbling fluidized beds, i.e. the sum of solid phase volume fraction at any moments is conserved or the mass of solid phase is conserved, a piDMD method that satisfies the mass conservation law is developed in present study. Fig.\ref{fig1} illustrates the concept of and the implementation procedure of piDMD for analyzing the coherent structures from the snapshots of solid volume fraction obtained from CFD-DEM simulation of gas-solid flows and then predicting the spatiotemporal dynamics of solid volume fraction.

The mathematical difference between DMD and piDMD is the calculation of the high-dimensional mapping matrix $\mathbf{A}$. It is clear that the bubbling bed in this study is characterized by the conservation of the sum of solid volume fractions, supposing that $\mathbf{z}_i$ represents the snapshot of solid volume fraction at the $i$th time, and the conservation of the sum of solid volume fraction is expressed mathematically as
\begin{equation}
{\left\| {{{\bf{z}}_{i + 1}}} \right\|_1} = {\left\| {{{\bf{z}}_{i}}} \right\|_1},
\label{A6}
\end{equation}
where the operator ${\left\| \cdot \right\|_1}$ denotes the 1-norm, and the 1-norm value of a vector is the sum of absolute value of each element of the vector. It is difficult to find a numerical solution for the coupling of the 1-norm with the DMD method.
But if you take the square root of each element in the snapshot vector and then square them, the resulting sum will be equal to the sum of the absolute values of each element in the vector.
Thus, the modified vector's 2-norm can be calculated with its elements set as the square root of each cell's solid concentration (i.e. $\left\| {{{\bf{z}}_{i,new}}} \right\|_2^2={\left\| {{{\bf{z}}_{i}}} \right\|_1}$).
This altered 2-norm (Eq.{\ref{A7}}) is equal to the original hydrodynamic solid concentration field's 1-norm (Eq.{\ref{A6}})
\begin{equation}
\left\| {{{\bf{z}}_{i + 1,new}}} \right\|_2^2 = \left\| {{{\bf{z}}_{i,new}}} \right\|_2^2.
\label{A7}
\end{equation}
\noindent Substitute Eq. \ref{A2} into Eq. \ref{A7} as

\begin{equation}
{\left\| {{\bf{A}}{{\bf{z}}_{i,new}}} \right\|_2^2 = \left\| {{{\bf{z}}_{i,new}}} \right\|_2^2.}
\label{A8}
\end{equation}

\noindent In matrix terminology, Eq.\ref{A8} holds if and only if $\mathbf{A}$ is unitary ($\mathbf{A}^*\mathbf{A}=\mathbf{I}$) \citep{Schonemann1966AGS}.
The feature that $\mathbf{A}$ is a unitary matrix ensures the physical law of mass conservation and serves as a constraint on the DMD regression, which constitutes the piDMD method in this study. Therefore, the piDMD regression is \citep{baddoo2023physics}
\begin{equation}
\arg \min {\left\| {{{\bf{Z}}_{1,new}} - {\bf{A}}{{\bf{Z}}_{0,new}}} \right\|_F},\,s.t.\,\mathbf{A}^*\mathbf{A}=\mathbf{I}.
\label{A9}
\end{equation}

The executable approach of piDMD-based prediction of solid volume fraction is summarized in Algorithm \ref{tab1}, steps $1$ and $5$ are exactly the same as DMD-based prediction, and steps $3$ and $4$ are the analytical solution process \citep{Schonemann1966AGS} for the piDMD regression. Predicting the solid volume fractions of the flow field at the next moment only requires repeating steps $5$ and $6$ in Algorithm \ref{tab1}.

\begin{table}
\normalsize
\renewcommand{\arraystretch}{1.25}{}
%\caption{The executable approach of piDMD-based prediction of solid phase volume fraction}
\captionsetup{labelfont={color=white}}
\caption{}
\centering
%\begin{center}
\begin{tabular*}{0.9\hsize}{l}
\hline
\multicolumn{1}{l}{\textbf{Algorithm 1}: piDMD-based prediction of solid volume fraction} \\ \hline
                     ${1.}$${\,}$Obtain two snapshot matrices $\mathbf{Z}_0$ and $\mathbf{Z}_1$ from Eq. \ref{A1}.                    \\
                     ${2.}$${\,}$Take the square root of each element of the matrix $\mathbf{Z}_0$ and $\mathbf{Z}_1$ to form ${\bf{Z}}_{0,new}$ and ${\bf{Z}}_{1,new}$.   \\
                     ${3.}$${\,}$Compute the SVD of ${{\bf{Z}}_{1,new}}{{\bf{Z}}_{0,new}^*}$,\\
\multicolumn{1}{c}{${{\bf{Z}}_{1,new}}{{\bf{Z}}_{0,new}^*} = {{\bf{U}}_{new}}{{\bf{\Sigma }}_{new}}{{\bf{V}}_{new}^*}$.}\\
                     ${4.}$${\,}$Calculate the high-dimensional mapping matrix $\mathbf{A}$ of the piDMD method with Eq. \ref{A9},\\
\multicolumn{1}{c}{${\bf{A}} = {{\bf{U}}_{new}}{{{\bf{V}}_{new}^*}}$.}\\
                     ${5.}$${\,}$Predict $\mathbf{z}_{j + 1, new}$ using ${\mathbf{z}_{j + 1, new}} = \mathbf{A} {\mathbf{z}_{j, new}}$.\\
                     ${6.}$${\,}$Take the square of each element of $\mathbf{z}_{j + 1, new}$ to obtain the solid volume fractions at the $j+1$th time, $\mathbf{z}_{j + 1}$.\\
                     \hline
\end{tabular*}
\label{tab1}
%\end{center}
\end{table}

\section{CFD-DEM method}\label{s3}
As mentioned earlier, the input data $\mathbf{Z}_0$ and $\mathbf{Z}_1$ for the DMD-based and piDMD-based analysis and prediction come from the computer simulation of a bubbling fluidized bed using CFD-DEM method.
However, it should be emphasized that the input data are not limited to CFD-DEM simulations, they can also be obtained from any other simulations where field variables of solid volume fraction are available, such as two-fluid model (TFM) \citep{gidaspow1994multiphase,wang2020continuum} and dynamic multiscale method \citep{chen2017dynamic,chen2018mesoscale}.
CFD-DEM method \citep{tsuji1993discrete,van2006multiscale} is one of the mainstream methods for numerical simulation of dense gas-solid systems. Table \ref{tab2} summarizes the main equations of the CFD-DEM method used in present study. Specifically, the volume-averaged Navier-Stokes equations of gas phase are solved in the Eulerian framework using the open-source software OpenFOAM$^\circledR$ \citep{holzmann2016mathematics}; the Newton's second law is applied in the Lagrangian framework to obtain the movement of individual solid particle \citep{van2006multiscale,golshan2020review,kieckhefen2020possibilities},
which is solved using the in-house GPU-based DEMms (discrete element method for multiscale simulation) software \citep{xu2011quasi,lu2014emms,xu2022discrete}. This coupled software has successfully simulated the hydrodynamics, heat, mass transfer, and chemical reactions in various fluidized beds \citep{lu2014emms,lu2016computer,zhang2017assessment,xu2019virtual,zhang2019experimental,lan2020long,liu2020long,
zhang2020cfd,zhao2020cfd,zhao2020computational,zhao2022euler,zhao2022cartesian,lu2023optimization}.
Therefore, the CFD-DEM data are directly used for piDMD and DMD analysis and as the benchmark for evaluating the short-term and long-term prediction of piDMD and DMD, without any further experimental validations.

CFD-DEM simulation of a typical bubbling fluidized bed is carried out, where the required physical parameters and numerical settings are reported in Table \ref{tab3}.
There are only air and solid particles in this bubbling fluidized bed, and its motion is such that air enters uniformly from the bottom of the bed and exits from the top, and the stationary piled-up spherical particles form a fluidized state under the action of air. During the period of steady fluidization, there are continuous generation, growth, coalescence and breakup of bubbles in the fluidized bed.

\begin{table}
\caption{Main equations of CFD-DEM method}
\label{tab2}
\begin{center}
\begin{tabular*}{0.9\hsize}{@{}@{\extracolsep{\fill}}l@{}}
\hline
Equations of translational motion of particle: \\
$m_a \frac{d^2 {\bf r}_{a}}{dt^2} = -V_a \nabla p + \frac{V_a \beta}{1 - \varepsilon _g}({\bf u}_{g} - {\bf v}_{a}) + {m_a}{\bf g}$ \\
                   \,\qquad \qquad $ + \sum_{b \in \text {contactlist }}\left(\mathbf{F}_{b a, n} + \mathbf{F}_{b a, t}\right)$ \\
Equations of rotational motion of particle: \\
$I_{a} \frac{d \omega_{a}}{d t}=\mathbf{T}_{a} \text {, where } I_{a} = \frac{2}{5} m_{a} R_{a}^{2} $\\
Torque of particle: \\
$\mathbf{T}_{a}=\sum_{b \in \text {contactlist }}\left(R_{a} \mathbf{n}_{a b} \times \mathbf{F}_{b a, t}\right) $ \\
Normal contact force between two particles: \\
$\mathbf{F}_{b a, n}=-k_{n} \delta_{n} \mathbf{n}_{a b}-\eta_{n} \mathbf{v}_{b a, n}, $ \\
where $\delta_{n}=d_{p}-\left|\mathbf{r}_{b}-\mathbf{r}_{a}\right|$, $\eta_{n}=\frac{-2 \operatorname{\ln e} \sqrt{k_{n} m^{*}}}{\sqrt{(\pi)^{2}+(\ln e)^{2}}} $\\
Tangential contact force between two particles:\\
$\mathbf{F}_{b a, t}=\left\{\begin{array}{cc}
-\eta_t \mathbf{v}_{b a, t} & \left|\mathbf{F}_{b a, t}\right| \leqslant \mu_f\left|\mathbf{F}_{b a, n}\right| \\
-\mu_f\left|\mathbf{F}_{b a, n}\right| \mathbf{t}_{b a} & \left|\mathbf{F}_{b a, t}\right|>\mu_f\left|\mathbf{F}_{b a, n}\right|
\end{array}\right.$\\
where $\mathbf{v}_{b a, t}=\mathbf{v}_{b a}-\mathbf{v}_{b a, n}, \eta_t=\eta_n, \mathbf{t}_{b a}=\frac{\mathbf{v}_{b a, t}}{\left|\mathbf{v}_{b a, t}\right|}$\\
Solid phase volume fraction of grid A: \\
${\varepsilon_{\mathrm{p}, \mathrm{A}} = \frac{{\sum\nolimits_{\mathrm{p}} {{{\mathrm{V}}_{\mathrm{p}}}} }}{{{{\mathrm{V}}_{\mathrm{A}}}}}}.$\\
Grid-averaged particle velocity of grid $A$:\\
$\mathbf{u}_{\mathrm{p}, \mathrm{A}}=\left\{\begin{array}{l}
\frac{\sum_{\mathrm{p}}\left(V_{\mathrm{p}} m_{\mathrm{p}} \mathbf{u}_{\mathrm{p}} g\left(\left|\mathbf{r}_{\mathrm{p}}-\mathbf{x}_{\mathrm{A}}\right| / h\right)\right)}{\sum_{\mathrm{p}}\left(V_{\mathrm{p}} m_{\mathrm{p}} g\left(\left|\mathbf{r}_{\mathrm{p}}-\mathbf{x}_{\mathrm{A}}\right| / h\right)\right)}, V_{\mathrm{A}}<v_{\mathrm{cs}}^{3} \\
\frac{\sum_{\mathrm{p}} m_{\mathrm{p}} V_{\mathrm{p}}}{m_{\mathrm{p}}}, V_{\mathrm{A}} \geq v_{\mathrm{cs}}^{3}
\end{array}\right.$\\
Gas-solid drag coefficient of grid A:\\
$K_{\mathrm{gp}, \mathrm{A}}=\left\{\begin{array}{l}
\frac{\sum_{\mathrm{p}}\left(\beta V_{\mathrm{p}}\left|\mathbf{u}_{g,\mathrm{A}}-\mathbf{v}_{\mathrm{p}}\right| g\left(\left|\mathbf{r}_{\mathrm{p}}-\mathbf{x}_{\mathrm{A}}\right| / h\right) / \varepsilon_{\mathrm{p}, \mathrm{A}}\right)}{\left|\mathbf{u}_{g, \mathrm{A}}-\mathbf{v}_{\mathrm{p}}\right| \sum_{c=1}^{N_{\mathrm{grid}}}\left(g\left(\left|\mathbf{x}_{c}-\mathbf{x}_{\mathrm{A}}\right| / h\right)\right)}, V_{\mathrm{A}}<v_{\mathrm{cs}}^{3} \\
\frac{\sum_{\mathrm{p}}\left(\beta V_{\mathrm{p}}\left|\mathbf{u}_{\mathrm{g}, \mathrm{A}}-\mathbf{v}_{\mathrm{p}}\right| / \varepsilon_{\mathrm{p}, \mathrm{A}}\right)}{\left|\mathbf{u}_{\mathrm{g}, \mathrm{A}}-\mathbf{v}_{\mathrm{p}}\right| V_{\mathrm{A}}}, V_{\mathrm{A}} \geq v_{\mathrm{cs}}^{3}
\end{array}\right.$\\
Gas phase mass conservation equation: \\
$\frac{\partial{(\varepsilon _g \rho _g)}}{\partial t}+\nabla \cdot (\varepsilon _g \rho _g{\bf u}_g) = 0$  \\
Gas phase momentum conservation equation: \\
$\frac{\partial{(\varepsilon _g \rho _g {\bf u}_g)}}{\partial t}+\nabla \cdot (\varepsilon _g \rho _g {\bf u}_g {\bf u}_g)= -\varepsilon _g \nabla p - {\bf S}_p $ \\
\; \quad \ \qquad \quad \quad \qquad \qquad \qquad $+ \nabla \cdot (\varepsilon _g {\bf {\tau}}_g)+\varepsilon _g \rho _g {\bf{g}}$   \\
Gas-solid drag force density: \\
${\bf S}_p = \frac{1}{V_{cell}} \sum_{a=1}^{N_{part}} \frac{\beta V_a}{1- \varepsilon_g} ({\bf u}_g - {\bf v}_a)\delta({\bf r}-{\bf r}_a) , $  \\
where Gidaspow \citep{gidaspow1994multiphase} drag correlation is \\
$\beta=\left\{\begin{array}{cc}\frac{3}{4} C_{D} \frac{\rho_{g} \varepsilon_{g} \varepsilon_{s}\left|t \mathbf{u}_{g}-\mathbf{v}_{a}\right|}{d_{p}} \varepsilon_{g}^{-2.65} & \varepsilon_{g}>0.8 \\ 150 \frac{\varepsilon_{s}^{2} \mu_{g}}{\varepsilon_{g} d_{p}^{2}}+1.75 \frac{\rho_{g} \varepsilon_{s}\left|t \mathbf{u}_{\mathrm{g}}-\mathbf{v}_{a}\right|}{d_{p}} & \varepsilon_{g} \leqslant 0.8\end{array}\right.$\\
with \\
$C_{D}=\left\{\begin{array}{cc}\frac{24}{R e}\left(1+0.15 R e^{0.687}\right) & R e<1000 \\ 0.44 & R e \geqslant 1000\end{array}\right.$\\
and \\
$Re=\frac{\varepsilon_g \rho_g d_p |{\bf u}_g - {\bf v}_a|}{\mu_g}$\\
Gas phase stress-strain tensor: \\
${\bf {\tau}}_g = \mu _g(\nabla {\bf u}_g + \nabla {\bf u}_g^T) - \frac{2}{3}\mu _g(\nabla\cdot{\bf u}_g){\bf I}$   \\
\hline
\end{tabular*}
\end{center}
\end{table}

\begin{table}
\small
\caption{Physical and numerical parameters used in CFD-DEM simulation}
\label{tab3}
\begin{center}
\begin{tabular*}{0.9\hsize}{@{}@{\extracolsep{\fill}}ll@{}}
\hline
\multicolumn{1}{l}{Parameter}&\multicolumn{1}{c}{Value} \\ \hline
\multicolumn{2}{c}{Gas phase}                        \\
                     Temperature, T (K)&  298                    \\
                     Dynamic viscosity, $\mu$ (Pa s) &  1.8$\times$$\/10^{-5}$         \\
                     Inlet superficial gas velocity, u (m/s)&  0.9                    \\
                     Minimum fluidization velocity, $u_{mf}$ (m/s) &0.3    \\
                     Molecular weight, M (kg/mol)&  $28.8\times10^{-3}$\\
                     Pressure, p (atm)&  1.0\\
                     Density, $\rho_g$ (kg/$\/m^3$)&  1.2   \\
                     CFD time step, dt (s) &1.0$\times10^{-5}$\\
\multicolumn{2}{c}{Particles}                        \\
                     Number of particle &167490                      \\
                     Diameter, $\/d_p$ (m)& $1.2\times10^{-3}  $                    \\
                     Density, $\rho_p$ (kg/$\/m^3$)&  1000   \\
                     Normal spring stiffness, $\/k_n$ (N/m) &32    \\
                     Tangential spring stiffness, $\/k_t$ (N/m) &32    \\
                     Friction coefficient&  0.1   \\
                     Restitution coefficient, $\/e_n,e_t$&  0.97   \\
                     Rolling friction coefficient& 0.01\\
                     Particle dynamics time step (s) &$\/1.0 \times {10^{{\rm{ - }}5}}$     \\
\multicolumn{2}{c}{Geometry}                        \\
                     Domain width, $x$ (m)&$0.1440  $                    \\
                     Domain thickness, $y$ (m)&$0.0192  $                     \\
                     Domain height, $z$ (m)&$0.3744$                   \\
                     Cell number, $\ {NX \times NY \times NZ}$ &$30 \times 4 \times 78$                     \\
                     Grid length in the x-direction, $\Delta x$ (m)&$0.0048$ \\
                     Grid length in the y-direction, $\Delta y$ (m)&$0.0048$ \\
                     Grid length in the z-direction, $\Delta z$ (m)&$0.0048$ \\
                     \hline
\end{tabular*}
\end{center}
\end{table}

\section{Result and discussion}\label{s4}
DMD and piDMD analysis and prediction are performed using the $5001$ time snapshots with its sampling frequency of $1000\,$Hz, corresponding to the CFD-DEM result from $5\,$s to $10\,$s. Analysis of the differential pressure drop of the whole bubbling fluidized bed indicates that the dominant fluctuation frequencies are $2\sim4\,$Hz (the results are not reported), furthermore, the sampling frequency of input data has a negligible effect on the prediction results provided that the data have contained sufficient information of bed dynamics, as shown in the Appendix. Therefore, only the results of a sampling frequency of $1000\,$Hz are reported here, which are analyzed and discussed in two parts: short-term prediction and long-term prediction.

\subsection{Short-term Prediction}\label{s5}
\begin{figure}
\centerline{\includegraphics[width=1\textwidth]{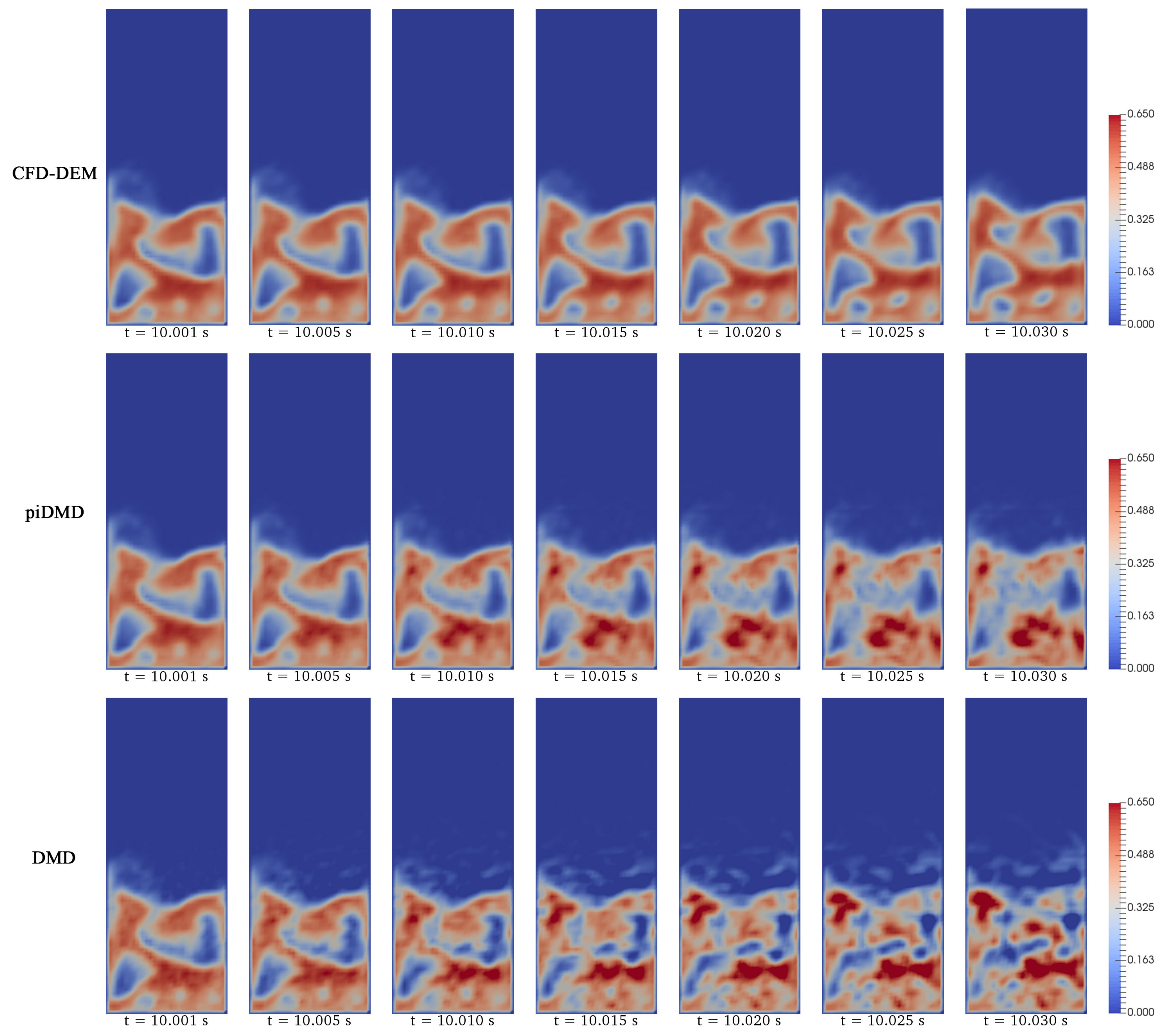}}
\caption{Snapshots of solid volume fraction obtained from CFD-DEM simulation, piDMD-based short-term prediction, and DMD-based short-term prediction.}
\label{fig2}
\end{figure}

\begin{figure}
\centerline{\includegraphics[width=1\textwidth]{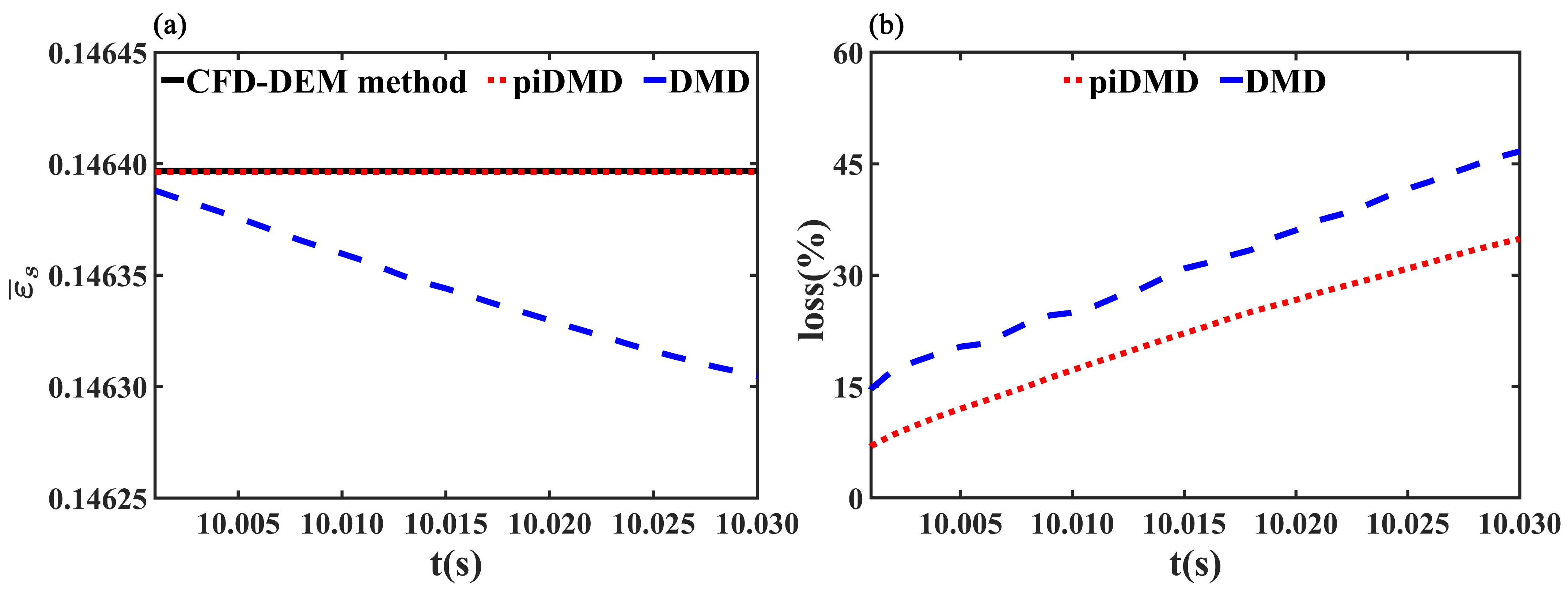}}
\caption{$(a)$ Simulation-domain-averaged solid phase volume fraction obtained from CFD-DEM simulation, piDMD-based short-term prediction, and DMD-based short-term prediction. $(b)$ The $loss$ obtained from piDMD-based short-term prediction and DMD-based short-term prediction.}
\label{fig3}
\end{figure}

Short-term prediction includes $30$ snapshots of solid volume fraction ${\varepsilon_s}$, corresponding to the time range from $10.001\,$s to $10.030\,$s. Fig.\ref{fig2} visualizes the results of piDMD-based and DMD-based short-term predictions at same moments. It can be seen that both DMD and piDMD can predict the solid volume fraction reasonably well, but the piDMD predicted flow field is closer to the CFD-DEM predicted flow field than those of DMD predicted flow field, qualitatively indicating that piDMD-based short-term prediction is more accurate. Fig.\ref{fig3} shows a quantitative comparison of piDMD-based and DMD-based short-term predictions. Fig.\ref{fig3}(a) illustrates the variation of simulation-domain-averaged solid phase volume fraction with time obtained from CFD-DEM simulation, piDMD-based short-term prediction, and DMD-based short-term prediction. The piDMD predicted and CFD-DEM simulated results are not only unchanged over time but also equal, whereas DMD predicted result follows a decreasing trend. The former observation indicates that the physical principle of mass conservation is faithfully satisfied, whereas the latter means the violation of the law of mass conservation since some mass is lost.
The ratio of the 2-norm of the deviation of the predicted snapshot from the real snapshot to the 2-norm of the original snapshot is used as a quantitative criterion to assess the difference between the predicted flow field and CFD-DEM simulated flow field, which is given by
\begin{equation}
loss = \frac{{{{\left\| {{{\bf{z}}_{\rm{p}}} - {{\bf{z}}_r}} \right\|}_2}}}{{{{\left\| {{{\bf{z}}_r}} \right\|}_2}}} \times 100\%  = \frac{{\sqrt {\sum\limits_{i = 1}^{i = N} {{{\left( {{z_{p,i}} - {z_{r,i}}} \right)}^2}} } }}{{\sqrt {\sum\limits_{i = 1}^{i = N} {{z_{r,i}^2}} } }} \times 100\%,
\label{A12}
\end{equation}
where $\mathbf{z}_p$ is the vector composed of the predicted snapshot at a given time using either DMD or piDMD, and $z_{p,i}$ is its element. Similarly, $\mathbf{z}_r$ is the vector composed of the real snapshot at a given time obtained from CFD-DEM simulation, and $z_{r,i}$ is its element. The smaller $loss$ value indicates that the predicted flow field is closer to the real one. Fig.\ref{fig3}(b) shows the $loss$ of the corresponding predicted flow field at each moment, and both curves show an increasing trend. However, the $loss$ of piDMD-based short-term prediction is smaller than that of DMD-based short-term prediction at each moment.

The results in this section suggest that the piDMD method is effective in achieving the mass conservation law, but the purely data-driven DMD method results in the violation of mass conservation law, which is a reason for why the results of piDMD-based prediction are consistently better than those of DMD-based prediction.

\subsection{Long-term Prediction}\label{s6}
\begin{figure}
\centerline{\includegraphics[width=1\textwidth]{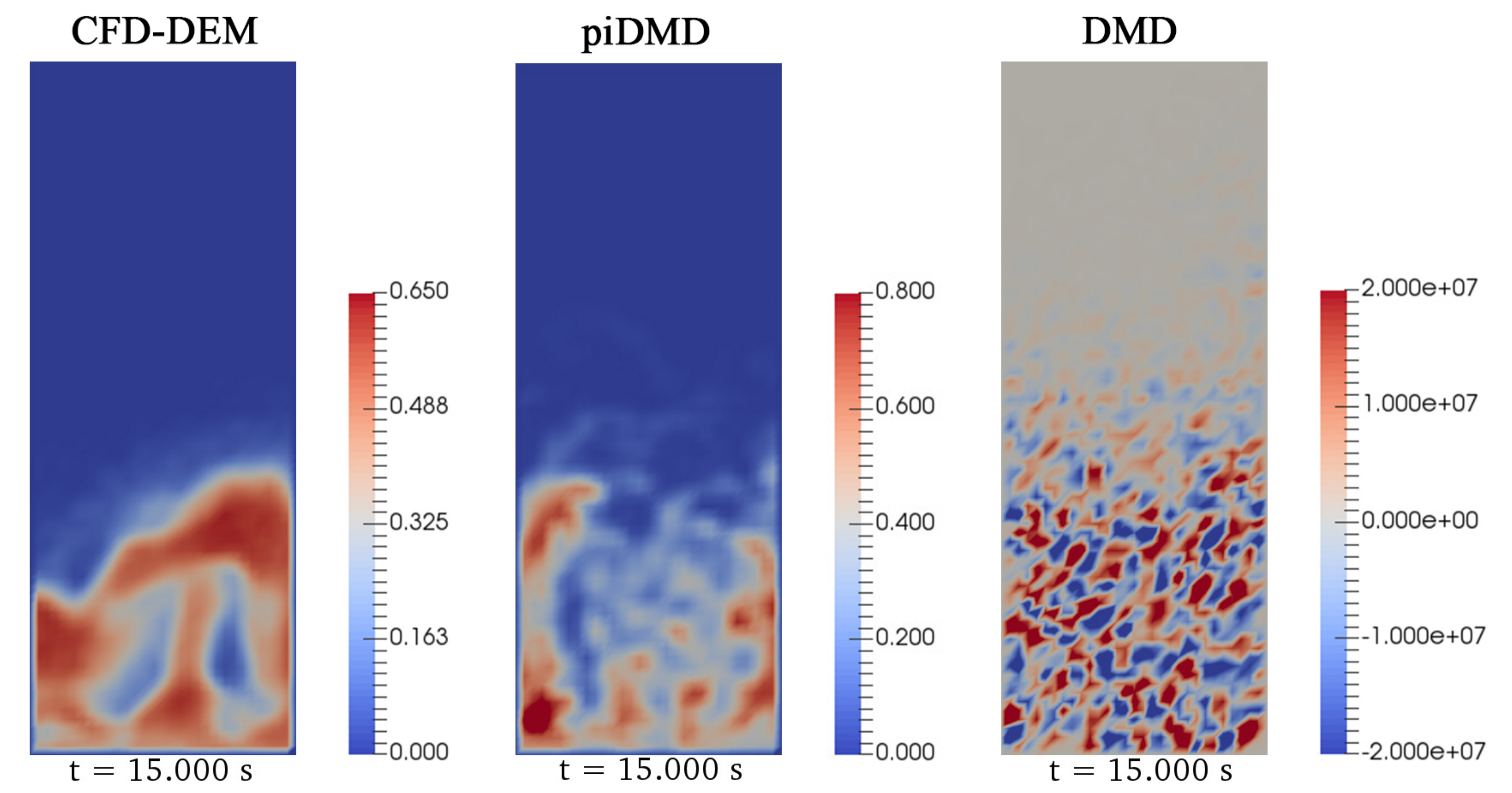}}
\caption{Snapshots of solid phase volume fraction at $t=15$ s, obtained from CFD-DEM simulation, piDMD-based prediction, and DMD-based prediction.}
\label{fig4}
\end{figure}

The time range for the long-term prediction is from $10\,$s to $50\,$s, and a total of $40001$ snapshots are predicted.
As shown in Fig.\ref{fig4}, by observing the difference in the color bar of CFD-DEM simulation, piDMD-based prediction, and DMD-based prediction at the physical time of $15\,$s, it is easy to find that the flow field predicted by the DMD method no longer possesses any physical reality of the flow field, but the flow field structure of piDMD-based prediction, although significantly different from that of CFD-DEM simulation, is still within the range of normal solid volume fraction values from a qualitative point of view.

\begin{figure}
\centerline{\includegraphics[width=1.02\textwidth]{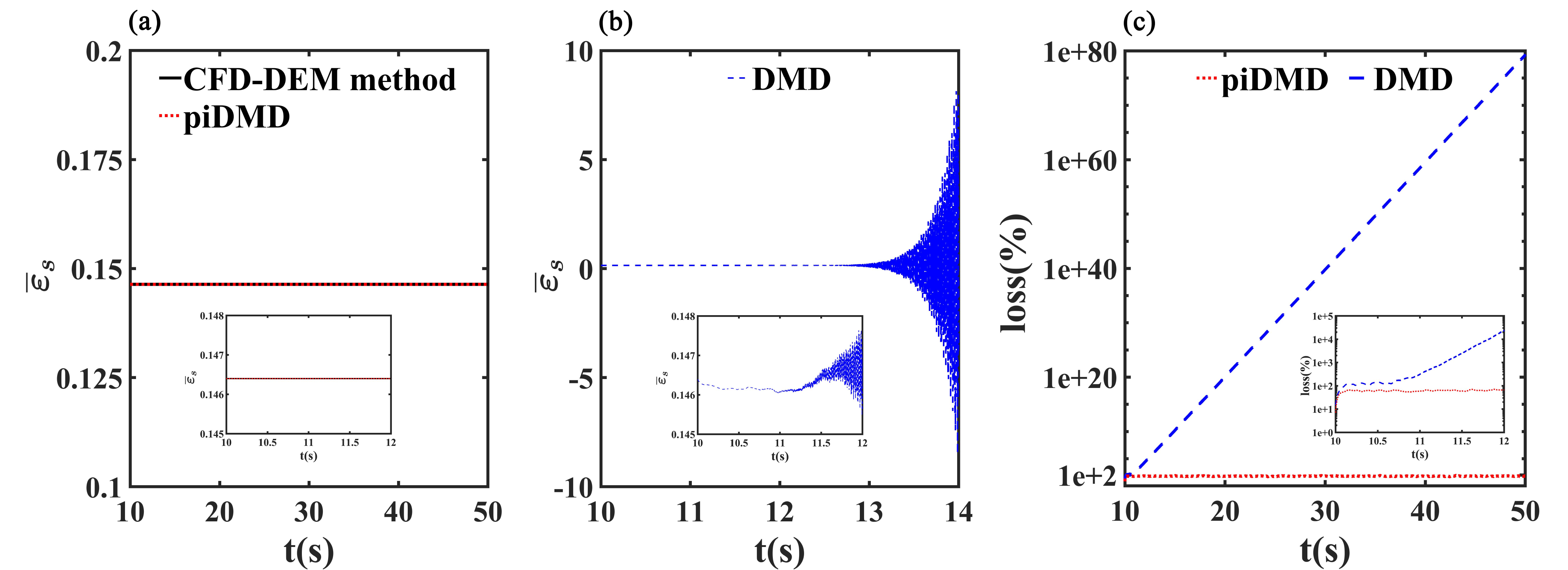}}
\caption{$(a)$ Simulation-domain-averaged solid phase volume fraction obtained from CFD-DEM simulation and piDMD-based long-term prediction. $(b)$ Simulation-domain-averaged solid phase volume fraction obtained from DMD-based long-term prediction. $(c)$ The $loss$ obtained from piDMD-based long-term prediction and DMD-based long-term prediction.}
\label{fig5}
\end{figure}

Fig.\ref{fig5}(a) and (b) show the variation of simulation-domain-averaged solid volume fraction with time for CFD-DEM simulation and piDMD-based long-term prediction, and DMD-based long-term prediction, respectively.
As can be seen from Fig.\ref{fig5}(a), the piDMD-based long-term prediction keeps the mean solid volume fraction unchanged, which is also consistent with the CFD-DEM result. Fig.\ref{fig5}(a) confirms that the piDMD predicted solid volume fraction strictly conforms to the physical law of mass conservation.
Nevertheless, it can be found from Fig.\ref{fig5}(b) that the mean solid volume fraction is not a fixed value, moreover, its fluctuation is increasing.
The numerical range of the curve in Fig.\ref{fig5}(b) is from $-10$ to $10$, which shows that the results of the DMD-based prediction are completely nonphysical, and the DMD method is unsuitable for long-term prediction of the hydrodynamics of bubbling fluidized beds.
Fig.\ref{fig5}(c) displays the loss of long-term prediction using the piDMD and DMD methods.
It is found that the $loss$ curve of piDMD-based long-term prediction increases in the early stage and then reach an asymptotic value, whereas the $loss$ of DMD-based long-term prediction increases rapidly with time.
From the subplot of Fig.\ref{fig5}(c), it can be noticed that the $loss$ value of piDMD-based prediction quickly stabilizes at a value around $62.5\,\%$.
The quantitative analysis shows that although the instantaneous results predicted by the piDMD method have errors, the errors do not magnify with the increase of prediction time, and the predicted flow field conforms to its inherent physical law (mass conservation). Clearly, how to reduce the $loss$ value is an interesting and critical issue for further improving the piDMD method and will be studied in future researches.

\begin{figure}
\centerline{\includegraphics[width=1.0\textwidth]{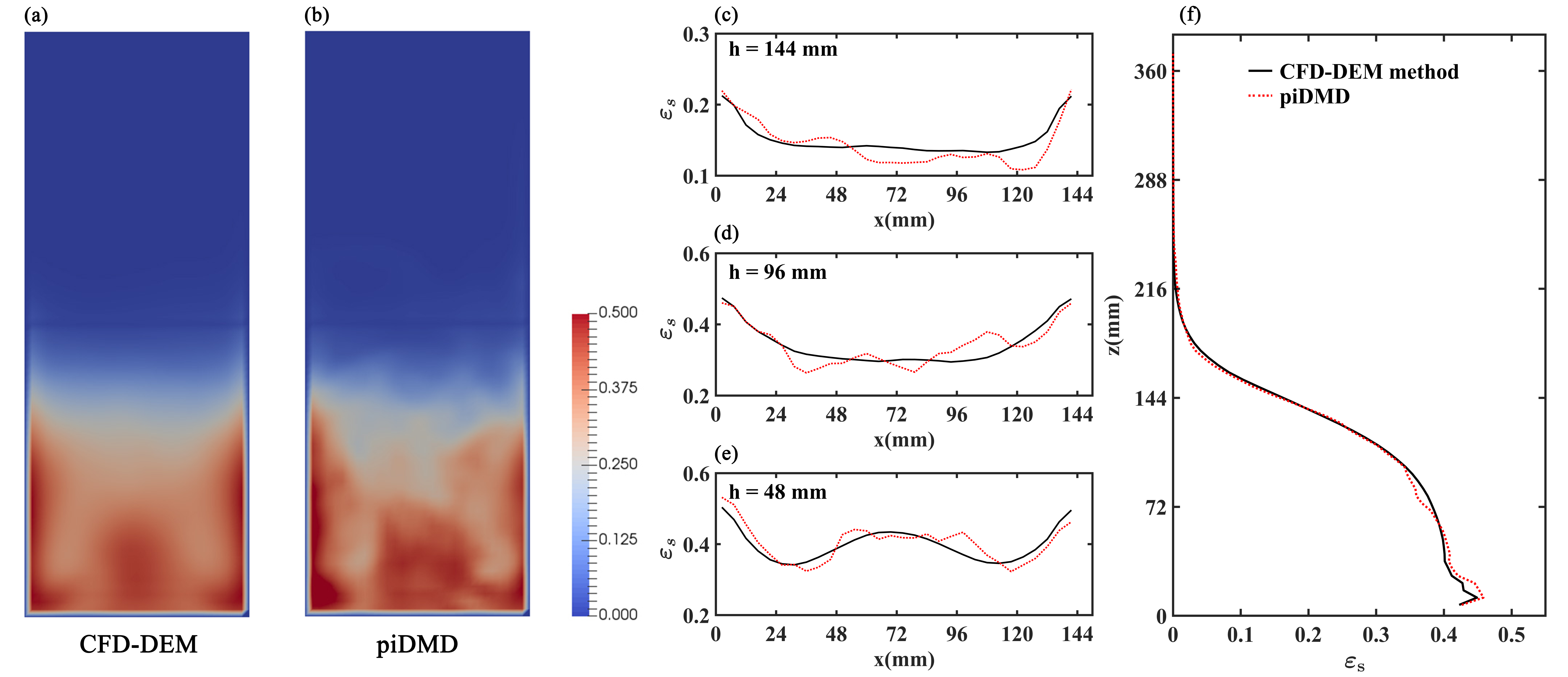}}
\caption{Comparison of time-averaged solid volume fractions obtained from CFD-DEM simulation and piDMD-based long-term prediction: $(a)$ and $(b)$ flow fields; the radial profiles at the bed heights of $(c)$ 144mm, $(d)$ 96mm, $(e)$ 48mm, and $(f)$ the axial profiles. }
\label{fig6}
\end{figure}

In addition to analyzing the accuracy of the instantaneous results of the piDMD-based long-term prediction, comparison between the time-averaged CFD-DEM results and the piDMD predicted results is shown in Fig.\ref{fig6}. Fig.\ref{fig6} $(a)$ and $(b)$ compare the time-averaged flow field from $10\,$s to $50\,$s predicted by piDMD method and simulated by CFD-DEM method, where the CFD-DEM simulation is the time-averaged result of the instantaneous flow field with a time interval of $0.001\,$s in order to consistent with the piDMD-based prediction, although the time step for CFD-DEM simulation is $10^{-5}\,$s. It can be observed that the structure of the time-averaged flow field of piDMD-based long-term prediction is similar to that of the time-averaged flow field simulated by CFD-DEM method. Fig.\ref{fig6} $(c)$, $(d)$, and $(e)$ show the radial profiles of the time-averaged solid volume fraction at the axial heights of $48\,$mm, $96\,$mm, and $144\,$mm, respectively.
Those figures reveal that the radial profiles of the time-averaged results of piDMD-based long-term prediction and CFD-DEM simulation are close at different heights. However, the results of piDMD-based long-term prediction are less smooth.
Fig.\ref{fig6}(f) displays the axial profiles of the time-averaged solid volume fraction obtained from CFD-DEM simulation and piDMD-based long-term prediction. Clearly, there are in a very good agreement.

Overall, the results of piDMD-based long-term prediction have satisfactory time-averaged results compared to CFD-DEM simulation, but the accuracy of the instantaneous results needs to be improved. As can be seen from ${\mathbf{z}_{j + 1, new}} = \mathbf{A} {\mathbf{z}_{j, new}}$, the error of each time step is accumulated to the next time step, which leads to the deviation of the instantaneous results, the development of physics-informed, streaming dynamic mode decomposition method \citep{hemati2014dynamic} might be a nice solution for further improving the prediction accuracy.
A simple comparison of the piDMD method and other prediction methods is in order,
the piDMD-based prediction is more efficient than the DeepVP proposed by \cite{qin2023deep}, which also predicted the voidage of bubbling fluidized bed, and took almost $525\,$h to build the DeepVP model at the first step, however, the whole process of the piDMD method predicts the physical time of $40\,$s flow field only takes $0.83\,$h. Furthermore,
the ability of the piDMD method to predict solid phase volume fraction over long periods is significantly better than the convolutional neural network (CNN) trained by \cite{bazai2021using}, whose model failed in generating correct contours of particle volume fraction in a fluidized bed compared to the result of CFD after $170$ time-steps (the time-step is $0.001\,$s).

\section{Conclusion}\label{s7}
A piDMD method is developed for predicting the solid volume fractions in a bubbling fluidized bed, where the physical law of mass conservation is integrated with the purely data-driven DMD method. On the basis of detailed comparison between the results of CFD-DEM simulation, piDMD-based prediction and DMD-based prediction, the following conclusions can be made:
(i) The piDMD-based prediction ensures that the predicted sum of solid volume fractions will not change at any time, following the physical law of mass conservation in the bubbling fluidized bed;
(ii) Both DMD and piDMD can predict the short-term behaviour of solid volume fraction reasonably well, but the piDMD method outperforms the DMD method in both qualitative and quantitative comparisons;
(iii) The piDMD method is able to predict the flow field in the long time, and the instantaneous results are not accurate enough compared with CFD-DEM simulation, but the $loss$ values of the predicted field are stable, and the time-averaged results are satisfactory in terms of radial and axial profiles of solid volume fraction;
(iv) The piDMD method is suitable for predicting the solid volume fraction distributions of bubbling fluidized beds very fast.

\section*{Acknowledgments}
This study is financially supported by the Strategic Priority Research Program of the Chinese Academy of Sciences (XDA29040200), the National Natural Science Foundation of China (11988102, 22378399), the Young Elite Scientists Sponsorship Program by CAST (2022QNRC001), and the Innovation Academy for Green Manufacture, Chinese Academy of Sciences (IAGM2022D02).

\section*{Appendix}
\begin{figure}
\centerline{\includegraphics[width=0.89\textwidth]{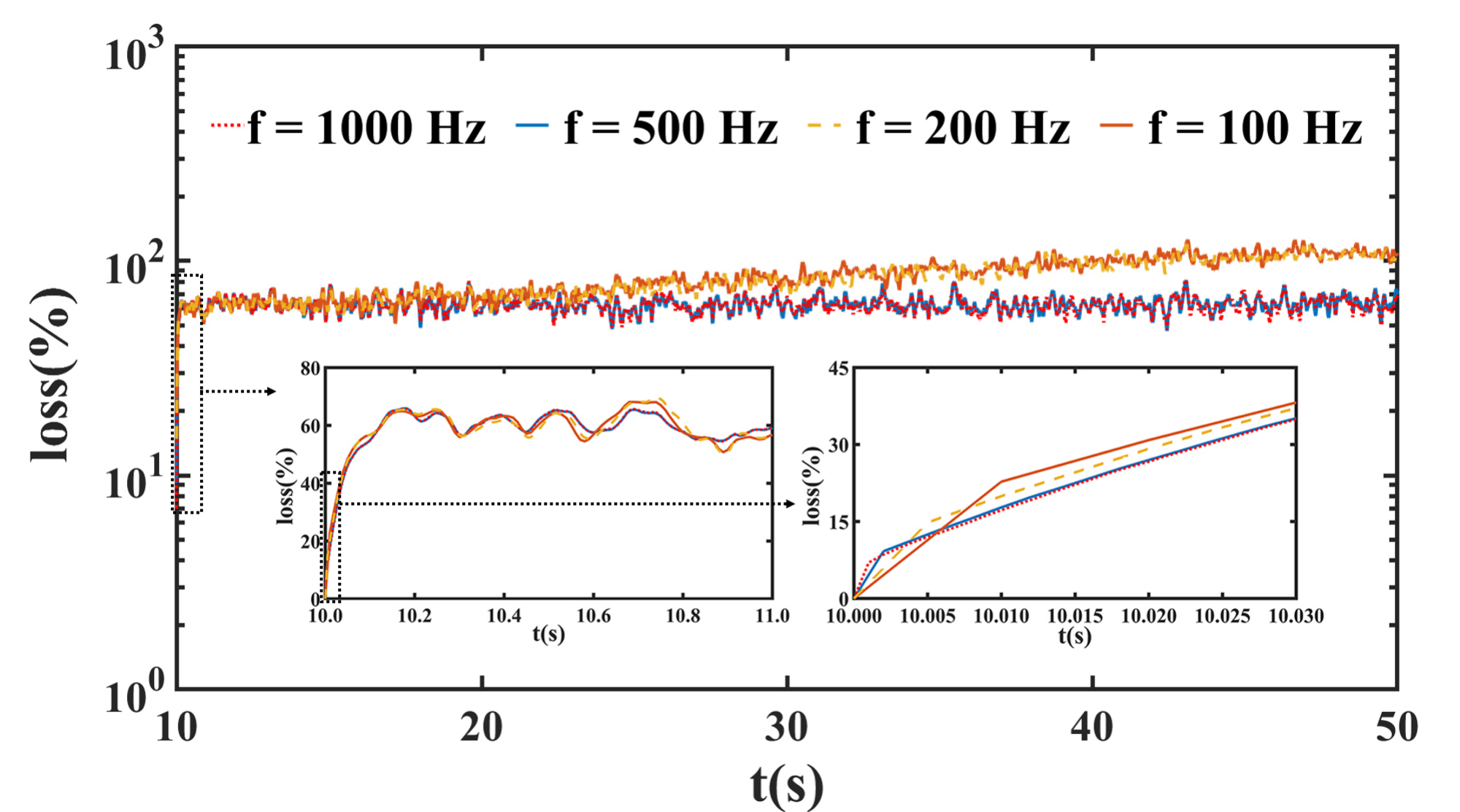}}
\caption{The $loss$ obtained from piDMD-based long-term prediction with its sampling frequency of $1000\,$Hz, $500\,$Hz, $200\,$Hz, and $100\,$Hz.}
\label{fig7}
\end{figure}

Fig.\ref{fig7} compares the $loss$ values of the piDMD-based prediction of the same sampling data length ($5\,$s to $10\,$s) and different sampling frequencies ($1000\,$Hz, $500\,$Hz, $200\,$Hz, and $100\,$Hz).
It is found that the loss values of the piDMD-based prediction with sampling frequencies of $1000\,$Hz and $500\,$Hz are the same, while the $loss$ values of the piDMD-based prediction with sampling frequencies of $200\,$Hz and $100\,$Hz are larger, indicating that the input data with smaller sampling frequencies do not contain enough characteristics of the bubbling fluidized bed.
%The piDMD-based prediction of the appropriate sampling frequency of $1000\,$Hz and $500\,$Hz, and the sampling frequency is the frequency of the predicted data. The larger the frequency of the same data length, the greater the number of data. This paper adopts the sampling frequency of $1000\,$Hz under the piDMD-based prediction in order to obtain more predicted data.

%\section*{References}
%\bibliographystyle{elsarticle-harv}
%\bibliographystyle{elsarticle-num}
%\bibliography{Reference}

\end{spacing}
\end{document}